# Room-Temperature Anomalous Hall Effect in Graphene in Interfacial Magnetic Proximity with EuO Grown by Topotactic Reduction.

Satakshi Pandey,[1] Simon Hettler,[2,3] Raul Arenal,[2,3,4] Corinne Bouillet,[1] Aditi Raman Moghe,[1] Stéphane Berciaud,[1] Jérôme Robert,[1] Jean-François Dayen,[1*] and David Halley[1*]

1-Université de Strasbourg, CNRS, Institut de Physique et Chimie des Matériaux de Strasbourg, UMR 7504, F-67000 Strasbourg, France

2-Laboratorio de Microscopias Avanzadas (LMA), Universidad de Zaragoza, 50018 Zaragoza, Spain

3-Instituto de Nanociencia y Materiales de Aragon (INMA), CSIC-Universidad de Zaragoza, 50018 Zaragoza, Spain

4-Araid Foundation, 50018 Zaragoza, Spain

## Abstract

**We show that thin layers of EuO, a ferromagnetic insulator, can be achieved by topotactic reduction under titanium of a $Eu_2O_3$ film deposited on top of a graphene template. The reduction process leads to the formation of a 7-nm thick EuO smooth layer, without noticeable structural changes in the underlying chemical vapor deposited (CVD) graphene. The obtained EuO films exhibit ferromagnetism, with a Curie temperature that decreases with the initially deposited $Eu_2O_3$ layer thickness. By adjusting the thickness of the $Eu_2O_3$ layer below 7 nm, we promote the formation of EuO at the very graphene interface: the EuO/graphene heterostructure demonstrates the anomalous Hall effect (AHE), which is a fingerprint of proximity-induced spin polarization in graphene. The AHE signal moreover persists above $T_c$ up to 350K due to a robust super-paramagnetic phase in EuO. This original high-temperature magnetic phase is attributed to magnetic polarons in EuO: we propose that the high strain in our EuO films grown on graphene stabilizes the magnetic polarons up to room temperature. This effect is different from the case of bulk EuO in which polarons vanish in the vicinity of the Curie temperature $T_c= 69K$.**

# I. Introduction

Combining the unique electronic properties of graphene and the spin degree of freedom of electrons is extremely attractive due to the expected large spin diffusion length of charge carriers in graphene [1,2,4], with pure spin current already demonstrated over several tens of micrometers [5,6]. Since pristine graphene is non-magnetic, exploiting graphene for spintronics requires either to inject spin-polarized carriers into graphene from ferromagnetic electrodes [3,7] or to induce a ferromagnetic phase in graphene that should lead to the spin polarization of carriers [8]. Several ways have been followed in order to achieve such a ferromagnetic graphene [9]. The proximity-induced ferromagnetism appears as a promising method: the hybridization of spin-polarized electrons in a ferromagnet, with the $\pi$-electrons of graphene should induce the exchange interaction necessary for long-range ferromagnetic interactions in graphene. This method involves ferromagnetic insulators, that do not shunt the current away from graphene.[10] Among the few available ferromagnetic insulators, YIG (Yttrium-Iron-Garnet) [11-13] EuS[14], EuO[15,16] and Fe-ferrites [17] have for example been studied in the recent years: Solis *et al.* [18] showed for instance through *ab-initio* simulations that YIG or EuO, used as local electrodes, could theoretically induce a large spin-polarization in graphene and even a subsequent giant-magneto-resistance (GMR) effect up to 100%, which is appealing for spintronics applications. This GMR effect has nevertheless to be experimentally demonstrated in graphene: fabricating such devices still requires the optimization of the ferromagnetic/graphene interfaces. For this purpose, spin-polarized transport properties in graphene are used as fingerprints of the proximity-induced spin polarization in graphene: anomalous Hall affect (AHE) [19], which is related to the appearance of magnetization in graphene, or non-local voltages revealing the Zeeman-spitting of graphene Dirac cones between up and down spins [13,14] are direct proofs of the spin polarization in graphene.

Among the insulating ferromagnetic that are candidate materials, EuO is a half-metal with a 1.1 eV gap [20]: the coupling between the *4f* $Eu^{2+}$ localized moments ($7\mu_B$) is archetypical of a Heisenberg ferromagnet [20], and shows a Curie temperature of 69K [20, 21,34]. As a noticeable drawback, EuO is not stable in an oxidizing environment (even in the air) [22,23] and turns into the $Eu_2O_3$ stable phase of the europium oxide, which is a non-magnetic insulator [24,25]. This makes the fabrication of EuO films relatively complicated. For this reason, and despite theoretical predictions of a large spin-polarization [18], only very few experimental works reported the fabrication of Gr/EuO films: Swartz *et al.* [15] or Klinkhammer *et al.* [16] used reactive molecular beam epitaxy (MBE) methods in an oxygen-limited regime called weak distillation mode to grow EuO on graphene – respectively on exfoliated graphene and CVD graphene on Ir(111)-, but they did not show AHE or other properties demonstrating the spin-polarization in graphene. More recently, Averyanov *et al.*, [26] used a similar MBE technique (again with an excess of Eu) on CVD graphene transferred on a $SiO_2$ substrate: some of the EuO films grown by this delicate method proved to be epitaxial (EuO [001] direction is along the normal to the plane) and turned out to induce an AHE effect in graphene up to 300K, well above the $T_c$ of bulk EuO, but exhibiting a surprisingly slow variation with the applied magnetic field, without clear saturation, at any temperature.

Besides studies on EuO/Gr systems, Mairoser *et al.* [23] showed, using an original topotactic technique, that thin layers of EuO could be performed on $YAlO_3$ substrates: for this purpose, they deposited the stable europium oxide - $Eu_2O_3$ - as a thin film by a sputtering method. The deposition of a titanium capping layer at high temperature -between 350°C and 550°C- led to a reduction of the $Eu_2O_3$ layer into the ferromagnetic EuO oxide (through the reaction $2Eu_2O_3 + Ti \rightarrow 4EuO + TiO_2$), which proved to exhibit good magnetic properties.

We show in this article the fabrication of a ferromagnetic EuO layer at the very graphene interface, without the need for sophisticated oxygen-assisted MBE techniques. This is made possible by combining the topotactic transformation of $Eu_2O_3$ layer grown on a graphene template, with the fine adjustment of the initial thickness of $Eu_2O_3$. Moreover, this method is demonstrated for the CVD graphene template [27], making our process relevant for applications and mass-scale studies. This relatively simple method could open the way to many applications for graphene spintronics. Moreover, the behaviour of the EuO magnetization in layers grown on graphene in this topotactic way is deeply modified, relatively to bulk EuO: we observe a robust super-paramagnetic phase up to 350K well above the bulk values for this phase. This suggests the possibility of exploiting the ferromagnetic EuO at room temperature, circumventing the usual limitation due to a low Curie temperature.

In these works, we focus on the thickness-dependent magnetic properties of the ferromagnetic EuO layers on top of a graphene CVD monolayer. Our goal is to precisely understand the direction of the magnetization, its localization in the oxide layer, and to have a quantitative estimation of its temperature dependency. Besides SQUID (superconducting quantum interference device) techniques, we use AHE in graphene as a probe of the magnetization of EuO at the interface, but we will mostly remain in a relatively "high temperature" range up to 350K; we therefore do not detail the specific proximity-induced electronic properties of graphene at low temperature (close to 2K) that are gate-voltage dependent: this will be extensively studied in a forthcoming paper.

## II-Topotactic growth of EuO and structural characterization

We use as growth templates commercially available samples (Graphenea) made of monolayer CVD graphene transferred on $SiO_2$/Si substrate. In order to get the highest surface quality for the EuO growth and to get rid of surface contamination such as polymer residues resulting from the CVD transfer process, the samples were first out-gassed in ultra-high vacuum for several hours at different temperature steps up to 400°C in a $10^{-10}$ mBar base pressure. The deposition of $Eu_2O_3$ was performed at room temperature using an e-beam evaporation gun in a Molecular Beam Epitaxy chamber, monitored by a quartz microbalance. The $Eu_2O_3$ evaporation leads to an increase of the oxygen partial pressure close $10^{-8}$ mBar. After deposition, the samples were heated to 450°C and then capped by a titanium layer in the same chamber (Fig.1(a)). After deposition of the capping layer, the samples were left 40' at 450°C to promote the reduction process by titanium of $Eu_2O_3$ into EuO before cooling down to room temperature.

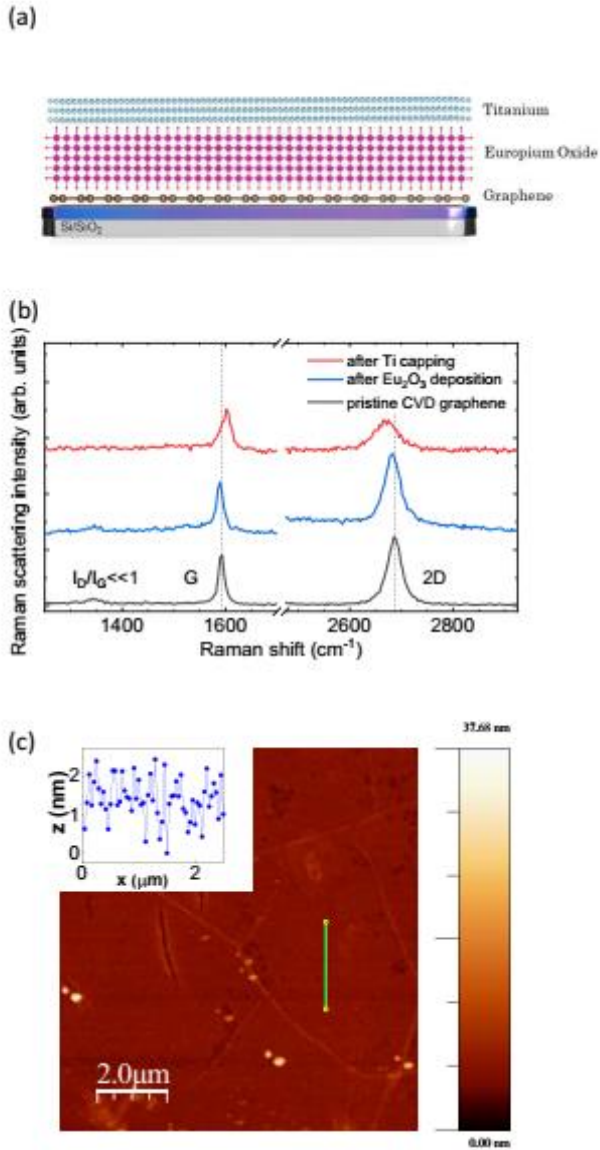

*Fig. 1: (a) Sketch of the deposited stack. (b) Raman spectra measured on the same sample at different steps of the growth: pristine CVD graphene on SiO₂, graphene covered by Eu₂O₃ at room temperature, graphene after Ti capping at 450°C. The vertical dashed lines are guides to the eye showing the positions of the G- and 2D- mode features in pristine graphene and illustrating the upshift of the G-mode and downshift of the 2D mode that develops after Ti deposition and which can be understood as a signature of significant electron doping. (c) AFM observation of the sample surface for $t_{Eu2O3}$ = 14nm. A typical height profile is given as inset. Notice that we observe typical wrinkles of CVD graphene domains.*

The impact on graphene of topotactic growth conditions is studied in Figs. 1(b): the Raman spectra on graphene are given for comparison on pristine graphene, after Eu₂O₃ deposition and after Ti capping at 450°C. The position and relative intensities of the characteristic D and G Raman peaks of graphene are not strongly modified and, noticeably, the $I_D/I_G$ ratio remains low at all steps, typically near or below 10%, confirming that the annealing of graphene below europium oxide is not detrimental to the quality of the graphene layer. Moreover, the observation of the titanium oxide surface by atomic force microscopy (AFM) in Fig.1 (c) proves a smooth surface of the layers: whatever the deposited $Eu_2O_3$ thickness, we observed continuous films covering the whole graphene layer with typical roughness below 0.5 nm.

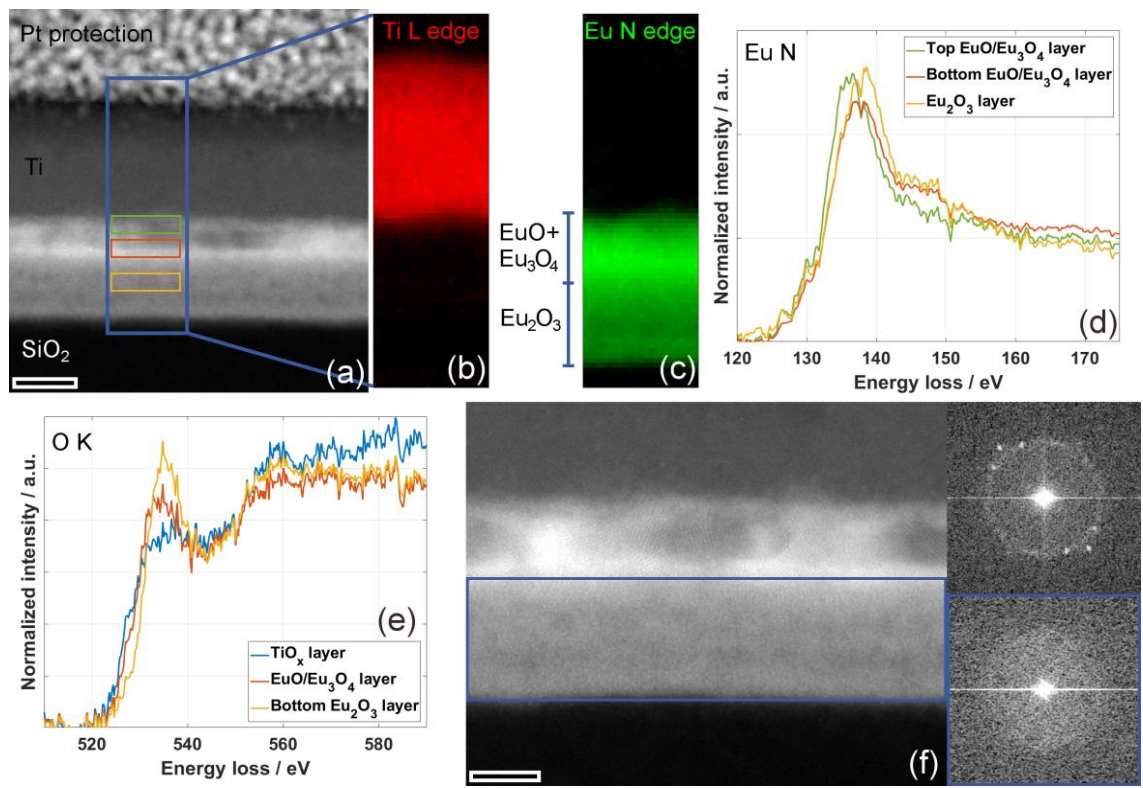

*Fig. 2: Electron microscopy observation on a 14-nm thick $Eu_2O_3$ sample on $SiO_2$, without graphene, capped by a titanium layer. (a) STEM cross section showing two different smooth layers in the europium oxide film, including darker clusters in the top layer. (b) and (c) EELS mapping obtained from the Ti L-edge and Eu N edge on the area shown in (a). No noticeable interdiffusion between Eu and Ti is observed. A difference in Eu edge intensity is observed between the top and bottom layers in europium oxide associated to a change in electronic density and stoichiometry from EuO to $Eu_2O_3$. (d,e) Comparison of EEL spectra of Eu-N and O-K edges, respectively, obtained from different layers of the cross section. (f) High-resolution STEM observation of the europium oxide film: the bottom layer is amorphous (see the Fourier Transform in the lower inset) whereas the top layer is polycrystalline (top inset with reflections of EuO marked).*

Cross sections for scanning transmission electron microscopy (STEM) were obtained by focused ion beam (FIB) preparation. A thick reference sample ($t_{Eu2O3}$ =14 nm, and $t_{Ti}$= 15 nm), grown directly on $SiO_2$, without graphene, shows by high-angle annular dark field (HAADF)-STEM imaging (Fig. 2(a)) two layers with different mass densities: one, darker, at the bottom is about 5-7 nm thick, whereas the brighter top oxide layer, at the titanium interface, has a thickness in the 7-8 nm range, and exhibits slightly darker included clusters. Electron energy-loss spectroscopy (EELS) spectrum-images (Fig. 2 (b) and (c)) and spectra (Fig. 2 (e) and (g)) confirm the presence of two different layers in the europium oxide film, as well as a low interdiffusion between Ti and Eu. Moreover, we observe (Fig. 2(f)) that the top europium oxide layer, including the inserted clusters, is polycrystalline. A detailed analysis of this polycrystalline, clustered layer by high-resolution TEM (HRTEM) imaging (Suppl. Inf.) indicates the presence of a EuO matrix with $Eu_3O_4$ clusters: they most probably form during the annealing and reduction process due to the close Ti layer. The presence of $Eu_3O_4$ clusters in a EuO matrix was reported in the case of EuO growth on graphene by Averyanov *et al*. [27]; $Eu_3O_4$ even formed a continuous film on graphene, without EuO, in the case reported by Aboljadayel *et al.* [30]. Concerning the bottom europium oxide layer in our samples, it appears to be amorphous (bottom inset in Fig. 2(f)): too far away from the Ti capping layer, it is not modified by the reduction process, and probably remains in its initial state.

On a similar sample, but grown on CVD graphene, HAADF-STEM observations (Fig.3(a) and (b)) confirm the same features, *i.e* a crystalline top EuO layer that is about 7-8 nm thick and includes darker clusters, and a 5-6 nm thick amorphous bottom layer. Again, the observed lattice fringes are consistent with a EuO-$Eu_3O_4$ mixture in the top layer.

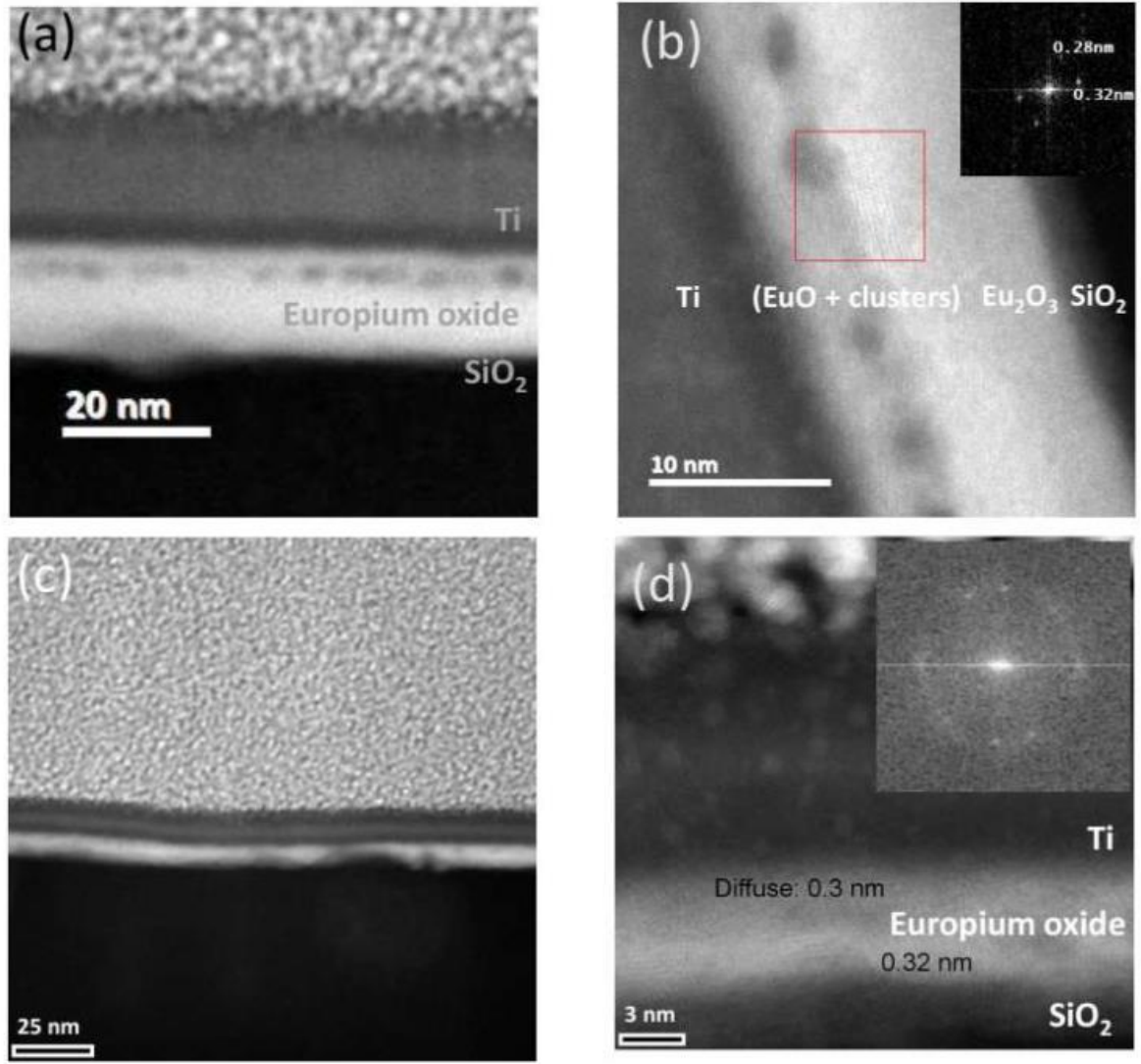

Fig. 3: Electron microscopy observation on cross sections of europium oxide samples on CVD graphene. (a) HAADF-STEM image of a 14-nm thick europium oxide film showing a 7-8 nm brighter top layer including darker clusters, and a 5-6 nm bottom layer. Notice the contrast in the 15-nm thick Ti layer, showing darker top and bottom layers, corresponding to oxidized titanium. (b) Zoom on the europium oxide layer: the top part, and the included clusters, are polycrystalline (see Fourier Transform in inset), the bottom part is amorphous. The lattice fringes (see inset) are consistent with the presence of $Eu_3O_4$ clusters and EuO. (c) HAADF-STEM image of a 7-nm thick europium oxide film, showing in some places fluctuations in the thickness, but no sign of a separation into two layers. (d) Zoom on the europium oxide film, which is polycrystalline down to the bottom interface and shows crystalline clusters in a diffuse matrix (the distance corresponding to clear reflections is given). This matrix again corresponds to $Eu_3O_4$ / EuO (see inset FFT). The graphene layer cannot be distinguished.

On a thinner sample ($t_{Eu2O3}$ = 7 nm) also grown on CVD graphene (Fig. 3(c) and (d)) we do not distinguish anymore the second amorphous layer at the bottom: $Eu_3O_4$ clusters are still observed in a nanocrystalline EuO matrix that now reaches the bottom interface – *i.e* the graphene layer-, but no $Eu_2O_3$ layer can be observed. Moreover, the europium oxide film, which remains globally smooth, exhibits in some places localized defects or constrictions in the oxide layer (Fig. 3(c)).

As a partial conclusion on TEM observations, we observe a low interdiffusion between titanium and europium oxide, as well as evidence of reduction of $Eu_2O_3$ into a 7-8 nm thick $EuO/Eu_3O_4$ layer. In samples with less than 7 nm $Eu_2O_3$ nominal thickness, the EuO layer appearing after the reduction process reaches the graphene bottom interface whereas in the case of thicker samples, the bottom layer, at the graphene interface remains in a $Eu_2O_3$ phase. These structural observations will be confirmed through magneto-transport measurements, as will be shown below.

## III. Magnetic measurements

Figure 4(a) shows the *M(H)* loops measured by SQUID under an in-plane magnetic field on a 14 nm-thick europium oxide reference sample directly deposited on a $SiO_2$ substrate, capped with 14 nm Ti: these curves confirm the reduction of $Eu_2O_3$ into EuO, leading to a large ferromagnetic signal. The coercive field $H_c$ equals 240 Oe and the remanence at *H=0* is close to 60 %. A magnetization of 2.7 Bohr magnetons per europium ion at 20K can be extracted from the SQUD measurement (assuming a uniform oxide layer on the whole 14nm europium oxide thickness). Taking into consideration the TEM observations of a 7-nm thick EuO layer on top of a non-magnetic $Eu_2O_3$ for such thick samples, we would obtain 5.4 $\mu_B$ per $Eu^{2+}$ ion, still lower than the 7 $\mu_B$ value expected in perfect EuO.

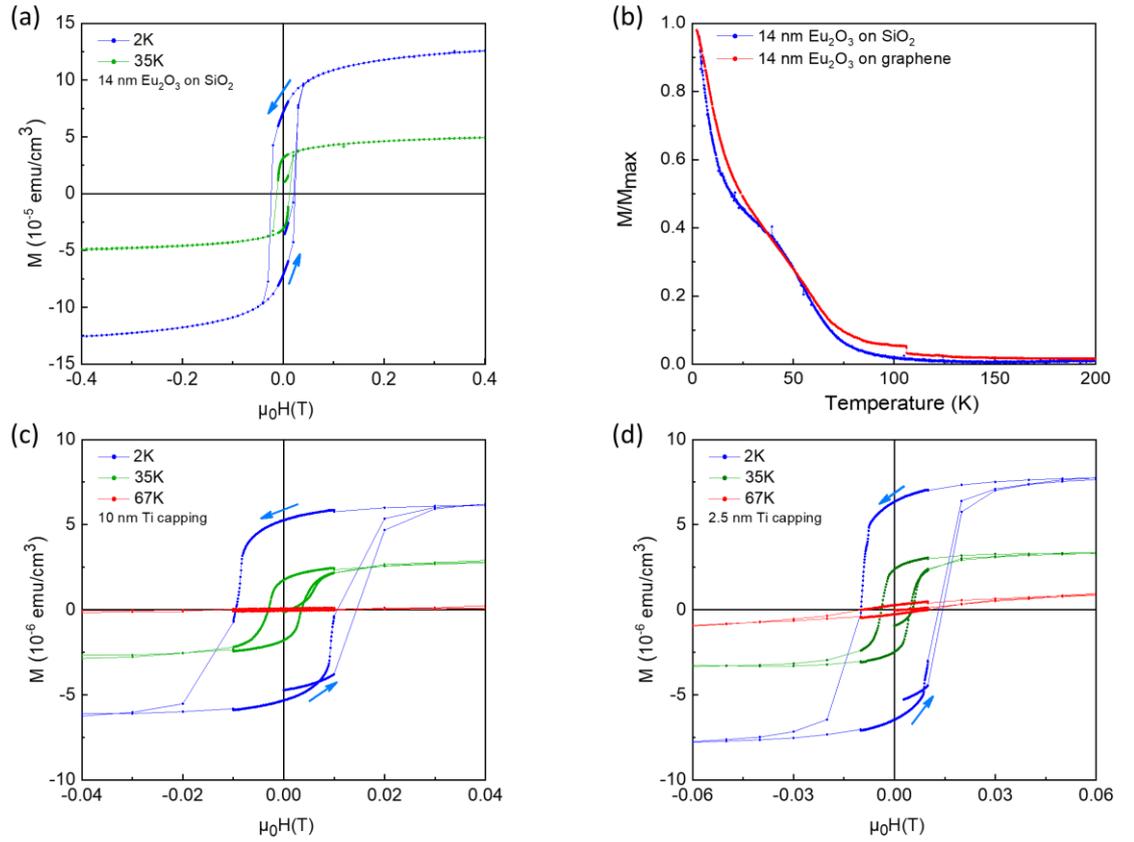

*Fig. 4 : In-plane SQUID measurements on "thick" samples with $t_{Eu2O3}$=14 nm, after Ti capping at 450°C: (a) M(H) loops at 2K and 35K on a sample directly deposited on SiO$_2$. (b) Normalized M(T) curves obtained with an applied magnetic field B=0.2T on samples grown on SiO$_2$ and on graphene. (c) and (d) M(H) loops on samples ($t_{Eu2O3}$=14 nm) grown on graphene with two different Ti capping thicknesses. The diamagnetic contribution of the substrate was subtracted.*

Temperature-dependent measurements of the magnetization under a magnetic field B=0.2 T (Fig. 4(b)) exhibit a decrease of *M* that drops to zero close to 100K. This value, higher than the Curie temperature *T$_c$ = 69K* in perfect bulk EuO crystals [20,21,34], is consistent with different works reporting an increase of *T$_c$* above 100K in non-perfectly stoichiometric EuO [33-35]. The shoulder in the *M(T)* curve of Fig. 4(b) between 70K and 100K is also in accordance with the observations made by Liu *et al* in the case of over-reduction of the EuO oxide [34] or La or Gd-doped EuO [36,37]. Moreover, a noticeable increase in magnetization is observed at very low temperatures (below 10K) which can be attributed to the contribution of Eu$_3$O$_4$ particles: Eu$_3$O$_4$ is a metamagnetic antiferromagnet [38] with *T$_N$* ≈*5.5K* in the bulk [39], which was shown to induce such a low-temperature magnetization when deposited on graphene layers [30]. This is consistent with our TEM observation of $Eu_3O_4$ clusters in the top oxide layer (Fig. 2 and 3). Furthermore, this explains that the evaluated magnetic dipole per europium ion does not reach the expected $7\mu_B$ in our samples, since a non-negligible fraction of Europium ions is included in this $Eu_3O_4$ phase.

As a comparison, we show in Fig. 4(c) the *M(H)* loops obtained for the same stack as in Fig.4(a) but deposited on a graphene CVD monolayer. The magnetization is strongly reduced in that case, but we are still dealing with EuO since the evolution that is observed (Fig. 4(b)) on *M(T)* curves is characteristic of the magnetic transition of EuO ($T_c$ is similar with and without graphene). If we assume, from TEM observations, a 7-nm thick reduced EuO layer, we obtain about 0.7 $\mu_B$ per $Eu^{2+}$ ion in the ferromagnetic layer. This moment is much below the expected 7 $\mu_B$ in pure EuO: this indicates that the fraction of europium ions entering the $Eu_3O_4$ phase might be larger in this sample, which drastically decreases the effective moment for all europium ions. This implies that graphene, perhaps through the strain induced in the oxide, or due to a modification of the structure (size, orientation of the grains), might influence the $Eu_2O_3/Eu_3O_4$ ratio forming under the titanium layer, by favouring the $Eu_3O_4$ phase, as already observed by Aboljadayel *et al.* [30].

Fig. 4(c) and (d) were obtained on similar samples of graphene *with $t_{Eu2O3}$=14 nm* but with two different titanium layer thicknesses, respectively of 10 nm and 2.5 nm. The *M(H)* curves are almost identical, with *$H_c \approx$ 100 Oe* and a high remanence at *H=0*. This similarity, whatever the titanium thickness, proves that down to 2.5 nm, there are enough titanium atoms in the capping layer to promote the reduction of $Eu_2O_3$. The following studied samples will thus be capped with a 2.5 nm Ti layer; this amount is sufficient to reduce the europium oxide layer, but low enough to avoid electrical shortcuts during transport measurements (at 2.5 nm, titanium is fully oxidized and thus insulating).

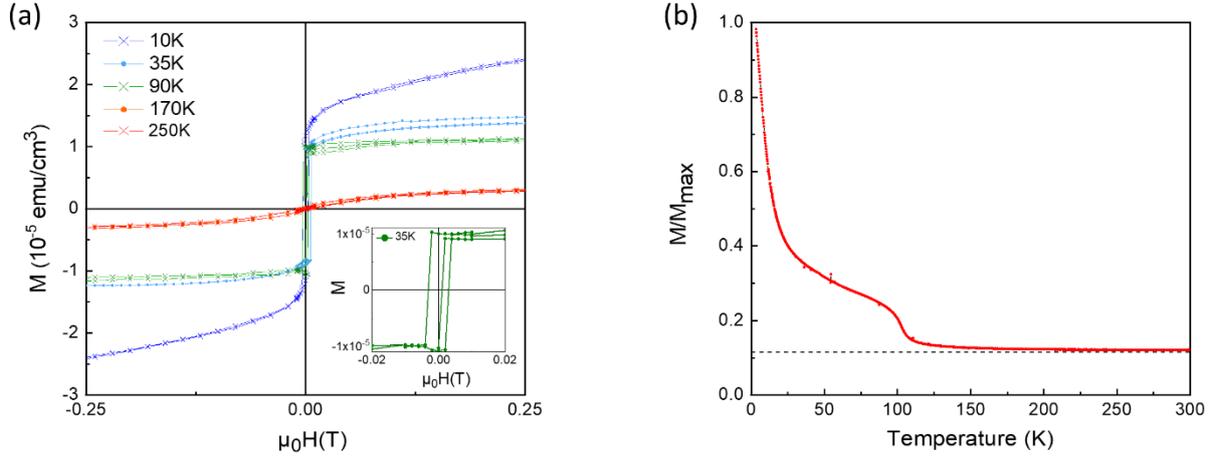

*Fig. 5: In-plane SQUID measurements on an intermediate thickness sample with $t_{Eu2O3}$=7 nm. (a) M(H) curves obtained at different temperatures. Inset: zoom on the magnetization reversal at low magnetic field at 35K. (b) Normalized M(T) curve obtained under B=0.2 T. The diamagnetic contribution of the substrate was subtracted. Notice that M does not drop to zero above $T_c$ but remains close to about one-tenth of $M_{max}$ (dotted line).*

By modifying the deposited europium oxide thickness $t_{Eu2O3}$, we observe that the decrease down to 7 nm leads to a smaller $H_c$ value (close to 25 Oe) (Fig. 5(a)), and lower remanence. Assuming a 7-nm EuO layer, we obtain from the magnetization value, about 1.5 $\mu_B$ per $Eu^{2+}$ ion. We observe on the *M(T)* curve (Fig. 5(b)) a $T_c$ value close to 100K, but, surprisingly, the magnetization value does not fall to zero above $T_c$: as confirmed by high-temperature *M(H)* curves, a non-zero signal survives up to room temperature, contrary to what was observed on thicker samples. This point will be elucidated below.

## IV-Magneto-transport measurements

We now explore the magneto-transport properties of our samples. Hall bars were fabricated by standard laser lithography techniques and physical graphene etching by oxygen plasma as detailed in previous works [31,32]. Gold electrodes were achieved by a lift-off process. Reference samples without graphene, deposited on $SiO_2$ substrates in the same conditions (*i.e.* capped at 450°C with titanium and annealed for 40') showed very large equivalent longitudinal $R_{xx}$ resistances, hardly measurable: we obtain for a $t_{Eu2O3}$=9 nm reference sample $R_{xx}$= *7.5 M$\Omega$* at 25K and *1.8 M$\Omega$* at 300K. This is more than 3 decades larger than the $R_{xx}$ values of graphene, making EuO conductivity negligible. This observation rules out a significant contribution from the oxide layer on the magneto-transport properties of the Gr/EuO heterostructure.

For thick film samples on graphene ($t_{Eu2O3}$= *14 nm*), the longitudinal $R_{xx}(H)$ curves (Fig.6(a)) are consistent with a classical magneto-resistance effect in graphene [41]. A negative magneto-resistance effect is moreover observed up to 20K, below *B=0.25T*, which can be attributed to a usual weak localization effect [41, 42] in graphene, as observed on the bare graphene reference sample (see inset in Fig. 6(a)). The $R_{xy}(H)$ curves (Fig. 6(b)) only show a simple linear behaviour. After subtracting the linear slope corresponding to the normal Hall effect, we only observe a non-significant noisy signal (inset in Fig. 6(b)). This confirms, as expected above, that there is no contact between graphene and EuO and thus no AHE related to the coupling between the ferromagnetic layer and graphene. This moreover proves that the magnetic stray field from the EuO top layer is not sufficient to induce parasitic Lorentz force within the graphene sensor.

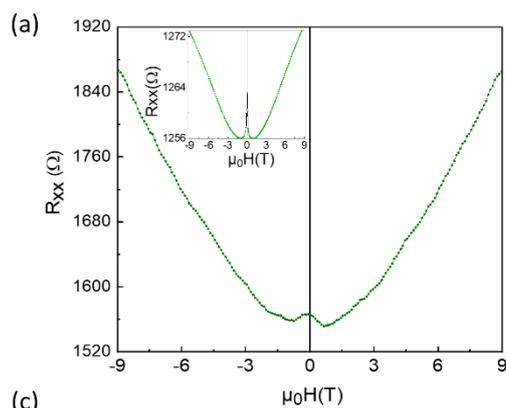
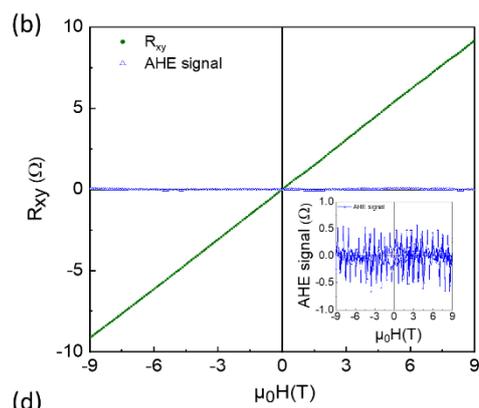
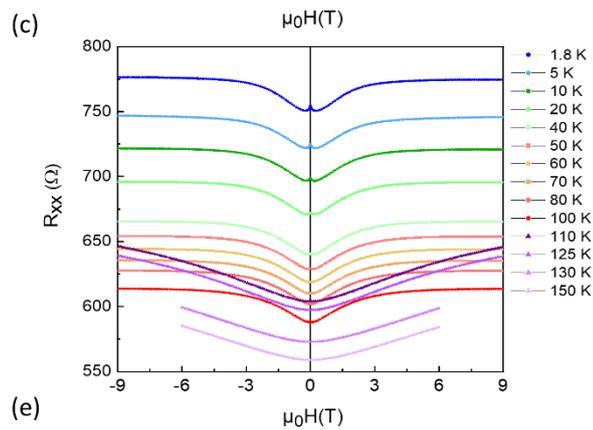
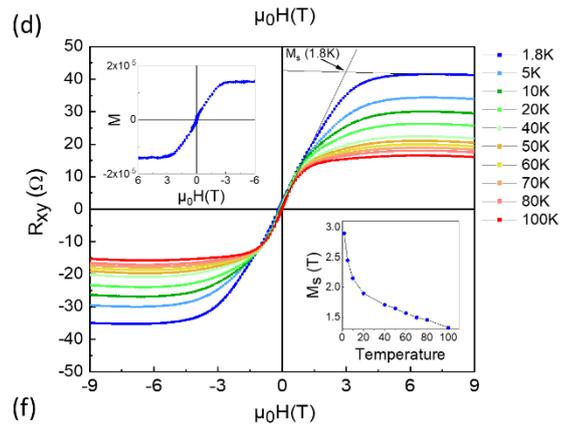
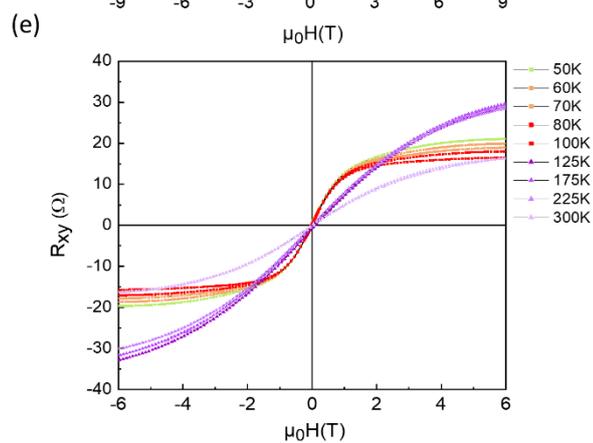
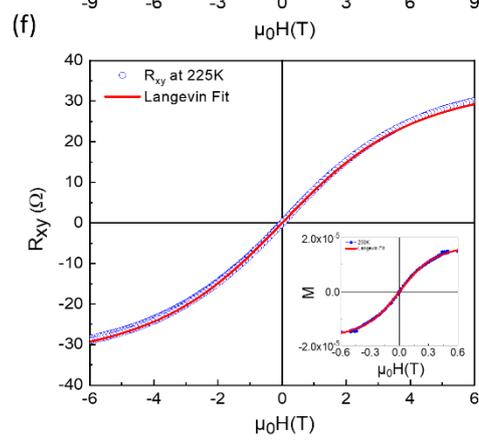

*Fig. 6 : . (a) and (b) Magneto-transport measurements at 25K on a "thick" sample with $t_{Eu2O3}$=14 nm grown on graphene. On the left panel $R_{xx}(H)$ longitudinal resistance of a Hall bar with H normal to the plane. Inset: same measurement on pristine graphene. On the right panel as-measured $R_{xy}(H)$ transverse resistance (red), and AHE signal after subtraction of a linear slope (blue). Inset: zoom on the zero-AHE signal exhibiting only noise. (c) and (d) Magneto-transport measurements on an intermediate-thickness sample with $t_{Eu2O3}$=7 nm. On the left panel $R_{xx}(H)$ curves at different temperatures showing below $T_c$ a saturation at high magnetic field. On the right panel as-measured $R_{xy}(H)$ curves at different temperatures below $T_c$ showing a saturation at high magnetic field and a constant slope around H=0. Upper inset: SQUID M(H) out-off plane measurement on a similar sample at 10K. Lower inset: $M_s$ plotted as a function of T.(e) same as in (d) with a zoom at lower field and extra curves measured above $T_c$ (purple). These high temperature curves exhibit a T-dependent slope. (f) $R_{xy}(H)$ curve at 225K with a fit by a Langevin function obtained with m= 205 $\mu_B$ Inset: M(H) curve measured by SQUID on a similar sample at T=225K, with a fit by a Langevin function obtained with m= 2350 $\mu_B$.*

As detailed previously, we have to decrease the thickness of the deposited europium oxide in order to promote the formation of EuO at the very interface with graphene. At a *thickness $t_{Eu2O3}$ = 9 nm*, we still do not observe any AHE signal (not shown), but at *$t_{Eu2O3}$= 7 nm* we observe the drastic effects shown in Fig. 6(d). The AHE part in the *$R_{xy}(H)$* curve dominates the normal Hall effect signal: indeed, the curves shown as-measured, without subtracting a linear slope. At 10 K for instance, they show a large variation in the low magnetic field range, with an almost linear behaviour, followed by saturation at about *B=2.2 T*, where the variation becomes much slower. As deduced from Figs. 4 and 5, the magnetization of the EuO layer lies spontaneously in-plane probably due to the demagnetizing field of the layer. During AHE measurements, it is forced by the perpendicular external magnetic field to align along the normal to the plane hard axis. The normal component of the magnetization $M_z$ is known [12] in that case to follow an almost linear behaviour relative to the applied magnetic field and to align along the normal to the plane at a field proportional to the saturation magnetization $M_s$ as $B = \mu_0 M_s$, which is required to overcome the demagnetizing field of the layer. That is further confirmed by SQUID measurements of $M_z$ (see inset on the left in Fig. 6(d)): The $R_{xy}(H)$ AHE signal and the $M_z(H)$ SQUID measurements follow nearly the same behaviour. This confirms their proportionality, with a saturation magnetic field value of the AHE signal that gives a rough estimation of the $M_s$ value. All those features clearly prove that we observe an AHE signal that is related to the $M_z$ component of the EuO layer: since this layer is directly at the interface with graphene, the proximity effect of EuO induces a spin-polarization of carriers in graphene. Notice that the observed saturation of the AHE signal differs from the results reported by Averyanov *et al.* [26] on EuO/Gr systems where the AHE signal evolved smoothly up to *B=9T*. Our results on the other hand are closer to the results shown by Wang *et al.* [12] in YIG/Gr systems, with an almost linear variation of the AHE signal with *H* followed by a clear saturation.

The longitudinal resistance $R_{xx}(H)$ is given on Fig. 6(c) at different temperatures. Beside weak localization effects at low magnetic field [40], a more surprisingly effect is observed: a positive magneto-resistive effect that saturates close to *B=2.2 T*, contrary to what we observe in bare graphene Hall bars, or in the case of thick -14 nm- europium oxide layers deposited on graphene where the parabolic-like magneto-resistance did not saturate with the applied magnetic field (Fig.6(a)). As the temperature is increased up to 100K, the shape of the curves remains globally the same, with a slight decrease of the magnetic field at saturation. We thus can draw a clear parallel between the $R_{xx}(H)$ and $R_{xy}(H)$ temperature-dependent behaviours: these observations show that the direction of the EuO magnetization directly influences the resistivity of the graphene layer. This clearly reminds theoretical predictions [43] of proximity-induced anisotropic magneto-resistance in *2D* materials at the interface with ferromagnetic layers; this effect dominates in our case the usual magneto-resistance observed in graphene. This again confirms that the behaviour of graphene is dramatically modified once in direct contact with the ferromagnetic EuO layer.

We now turn to the "high-temperature regime" above $T_c$: As can be seen on Fig. 6(d), the shape of $R_{xx}(H)$ curves is modified when *T* crosses a threshold above 100K which corresponds to $T_c$ as evaluated from SQUID measurements. The low magnetic field regime (roughly speaking, below *B=2.2 T*) disappears and we just observe a parabolic-like change of $R_{xx}(H)$ curves on the whole magnetic field range, with a larger amplitude in the variation of $R_{xx}$. In a correlated way, the $R_{xy}(H)$ curves (Fig. 6(e)) show a transition: the slope at low magnetic field changes with *T* -contrary to the case of the low temperature regime- and the curve' shape does no more follow a linear behaviour followed by saturation. Indeed, no clear magnetic saturation is observed anymore: as can be seen on Fig. 8(b), these curves follow a Langevin law, with *M(H)* given by $M(x) \propto \frac{1}{\tanh x} - \frac{1}{x}$ where $x = \frac{m H}{k T}$ and *m* is the average magnetic dipole carried by nanometric magnetic objects that will be defined below [44,45]. SQUID measurements carried out in this "high temperature regime" confirm a similar behaviour, with again good fits to the experimental data using Langevin functions (inset in Fig. 6(f)), nevertheless, with a larger value of *m* obtained from the adjustment of these curves. This dramatic change in the *M(H)* shape points to a transition from a ferromagnetic state to a super-paramagnetic state in the high temperature regime, at $T_c$. Although EuO is no more ferromagnetic, its magnetization survives above the Curie temperature in an assembly of small objects and still drives the AHE in graphene, which implies that there is still a spin-polarization of the carriers in graphene. Such a magnetic state has been reported in bulk EuO, but only in the vicinity of $T_C$, or in doped EuO, below 140K [34]. This phase was explained by the presence of magnetic polarons [34,44,45] in the oxide. These polarons couple antiferromagnetically to the localized spins on $Eu^{2+}$ ions and subsequently induce ferromagnetic exchange coupling between each other. Within the polaron radius, all ionic spins should thus contribute to a magnetic dipole *m*, yet with a non-negligible disorder [44].

This magnetic phase moreover explains the magnetic signal preserved over $T_C$ that was observed by SQUID in Fig. 5(b): contrary to magnetic polarons in bulk and doped EuO that vanish above 140K [34, 44], the polaron in strained EuO/graphene heterostructures are here preserved up to room temperature, (and even up to 350K, see Suppl. Fig.2).

Another proof of the proximity effect in graphene is given by gate-voltage dependant Anomalous Hall Effect measurements. By applying a gate voltage $V_g$ on graphene through the $SiO_2$ insulating layer we can modulate the magneto-transport properties of graphene on Hall bars: the transition from electron to hole carriers takes place between $V_g$=-85V and $V_g$=-100V as $V_g$ is decreased, corresponding to a change in the Hall effect sign, i.e. of the high magnetic field slope' sign of $R_{xy}(H)$ curves shown in Fig 7(a). The AHE signal shown in Fig. 7(b) on the other hand does not reverse its sign, as previously reported for Graphene magnetically proximitized with YIG [12] for example. This again proves that the AHE and normal Hall effect follow two different mechanisms: AHE is therefore not due to stray fields from EuO but to a proximity effect at the graphene interface.

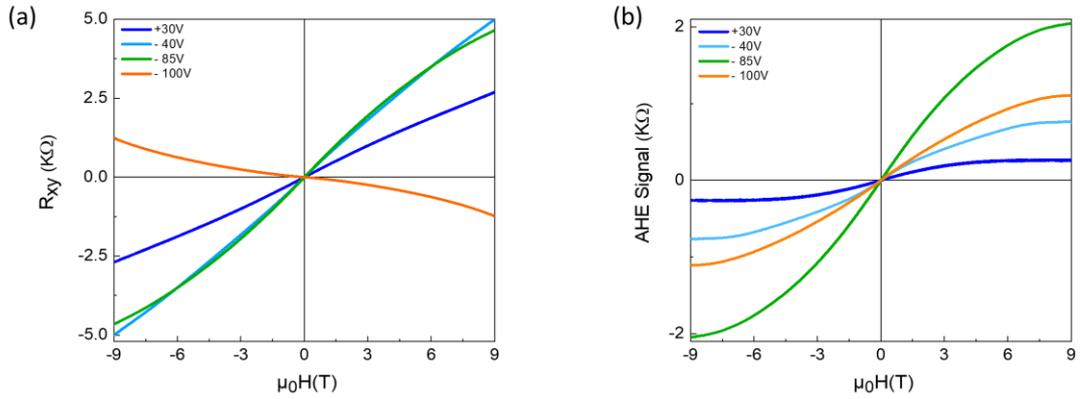

*Fig. 7: Hall measurements at different gate voltages observed at T=40K. (a) $R_{xy}$ measurements including the normal and anomalous Hall effect. (b) AHE signal after subtracting the slope corresponding to the normal Hall effect.*

If we now consider a thinner sample with 5 nm initial $Eu_2O_3$, we still observe (Fig. 8(a)) an AHE signal in graphene up to room temperature. This reveals the presence of a robust magnetization in EuO. From 100K to 300K all $R_{xy}(H)$ curves can indeed be adjusted to Langevin functions. The amplitude of the AHE signal decreases noticeably with increasing *T* whereas the shape of the curves only slightly evolves. The adjustments by Langevin functions in the magnetic polaron phase enable us to extract the *m* parameter [36,45] at different temperatures, as plotted in Fig. 8(b): *m* increases with *T* and amounts to *200$\mu_B$* at room temperature. A similar analysis (Fig. 8(b)), performed on the magnetic polaron phase of the 7-nm thick $Eu_2O_3$ sample of Fig. 6, proves a similar behaviour, with *m=300$\mu_B$* at room temperature. Finally, in a thinner sample with $t_{Eu2O3}$ = *4 nm*, we did not observe any magnetic transition into a ferromagnetic phase at low temperature: the magnetic polaron phase is observed on the whole [2K-300K] temperature range and leads to an almost linear *m(T)* dependency (Fig. 8(b)).

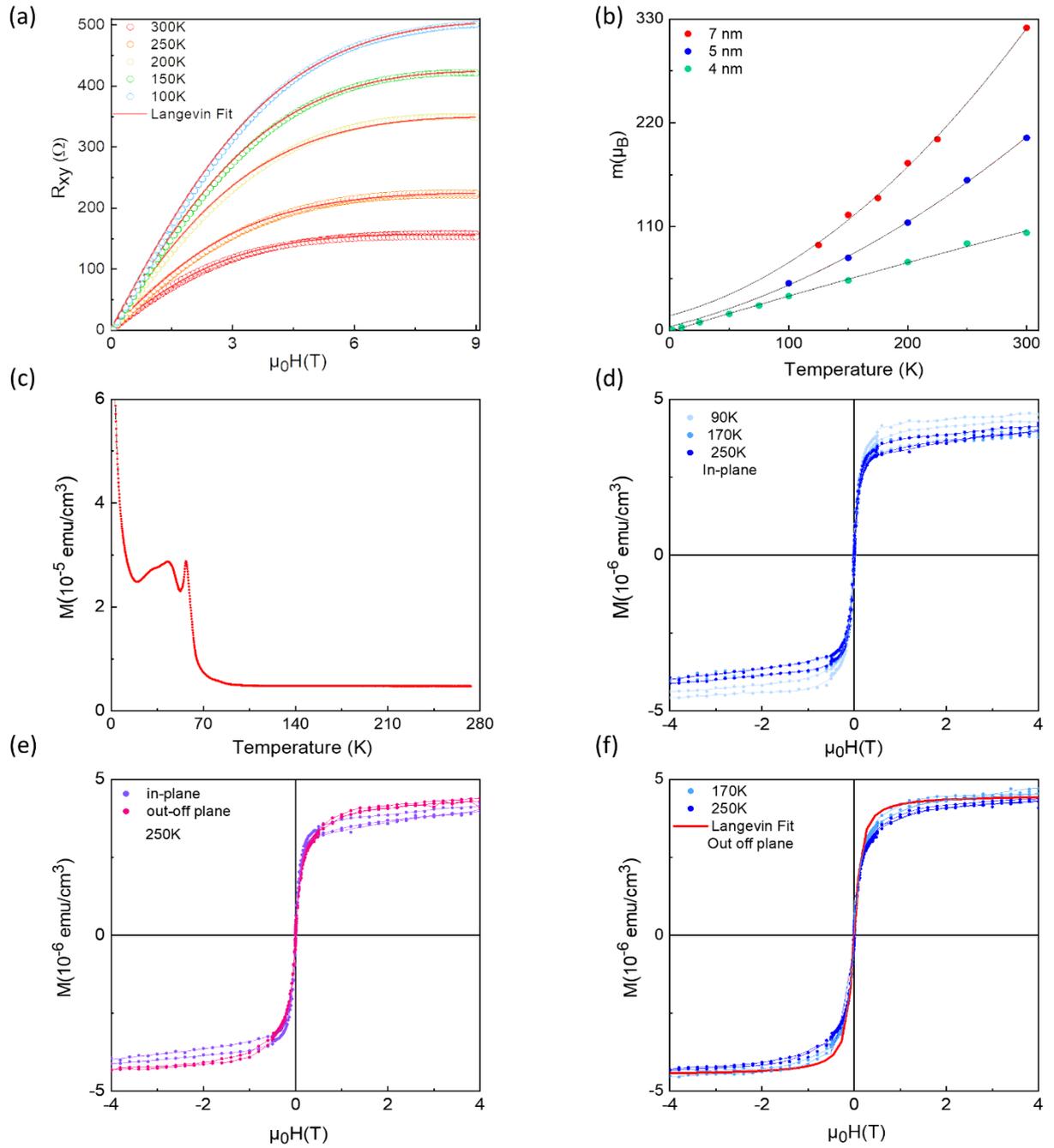

*Fig. 8: (a) AHE curves obtained under positive magnetic field on a "thin" sample with $t_{Eu2O3}$= 5 nm, at different temperatures above $T_c$. For each curve, a fit by a Langevin function is shown (red lines). (b) Plot of the m(T) parameters obtained above $T_c$ from fits of $R_{xy}(H)$ curves by Langevin functions, for three samples: (red) $t_{Eu2O3}$= 7nm, with $T_c \approx 110K$, (blue) $t_{Eu2O3}$= 5 nm, with $T_c \approx 62K$, and (green) $t_{Eu2O3}$= 4 nm, with no observable $T_c$ (the magnetic behaviour is superparamagnetic at any T for these thin layers). A second-order polynomial curve is given as a guide to the eyes for the three curves (black lines). (c)-(f) SQUID measurements on a "thin" sample with $t_{Eu2O3}$= 5 nm. The diamagnetic contribution of the substrate was subtracted: (c) Normalized M(T) curve. Again, M does not drop to zero above $T_c \approx 62K$. (d) In-plane M(H) loops at different temperatures above $T_c$. The curves almost superimpose whatever the temperature, up to 300K. (e) Comparison of in-plane and out-off plane curves at 250K. (f) Out-off-plane M(H) loops at different temperatures above $T_c$, that are almost superimposed. A fit by a Langevin functions is given.*

In addition to the SQUID observations shown in Figs. 5(a) and 6(f) on 7-nm tick samples, we confirmed the high-temperature magnetic phase by magnetometry measurements on our 5-nm thick (Fig. 8) sample: the in-plane *M(T)* loops (Fig. 8(c)) show a decrease of $T_c$ close to 50K relative to thicker films, and again a non-zero, almost constant, magnetization above $T_c$ up to room temperature. Moreover, the in-plane *M(H)* curves (Fig. 8(d)) above 50K are close to each other whatever *T*. Out-off plane measurements (Fig. 8(e)) exhibit a similar behaviour as a function of *T*, and *M(H)* loops are close to in-plane curves (Fig. 8(f)). This similarity is another proof of the assembly of small magnetic dipoles: in a continuously magnetic film, the demagnetizing field should systematically induce a difference between in-plane and out-off plane measurements: we are thus dealing with *3D* localized objects inside the thin film. The *M(H)* curves can here be adjusted by Langevin functions from which we extract *m* values that are much larger than the values obtained from AHE curves: at 300K *m* equals 6000 $\mu_B$, instead of 200 $\mu_B$ as evaluated from AHE. Notice that the same discrepancy was observed on the 7-nm thick sample (Fig. 6(f)), for which SQUID measurements gave *m* close 3500 $\mu_B$ at room temperature instead of about *300 $\mu_B$* from AHE.

# V-Discussion

The main question raised by our experimental results concerns the nature and origin of the robust polaron-induced SP phase.

An order of magnitude of the polarons' radius in the different samples can be obtained by supposing that polarons perfectly couple $Eu^{2+}$ spins carrying a $7\mu_B$ magnetic dipole; this is a rough approximation which neglects the spin-disorder inside polarons, which can be large in EuO [44]. We can, by this way, assess the diameter of $300\mu_B$-polarons to about 1.2 nm. As a comparison, in bulk samples, the radius of a polaron in EuO was reported [47] to vary with temperature, but a typical diameter was about 3 nm, which implies about 1000 europium ions, and thus *m= 7000 $\mu_B$* for the magnetic object associated to the polaron. The values obtained in our samples, from SQUID measurements (a few thousands of Bohr magnetons) and from AHE measurements (a few hundreds of $\mu_B$), are thus lower than observed in bulk EuO. Moreover, the increase of *m* with *T* (Fig. 8(b)) suggests either an increase of the polarons diameter or of the spin alinement inside polarons with temperature: indeed, those parameters were shown [45,47] to vary with the carrier density, which itself varies with *T*.

Concerning the origin of this SP phase, we have to underline that a similar phase was reported by Averyanov *et al.* [26] in the case of epitaxial EuO on graphene: the AHE signal revealed a SP phase at high temperature (up to 300K) but also in the low temperature regime (even at 2K) if we consider their smooth, non-saturating *M(T)* curves, with almost no temperature-induced change of the curves' shape. This corresponds to the behaviour of our thinner sample (4-nm thick) but differs from our thicker samples which exhibit a clear "usual" ferro-magnetic sate below $T_c$. Nevertheless, those results combined with our observations point to a strong influence of graphene which promotes this SP phase up to high temperature. Averyanov *et al.*, by analogy with the work of Katmis *et. al.*, [46] on $EuS/Bi_2Se_3$ systems, suggested that the polaron-induced SP phase in EuO on graphene could be driven by the large spin-orbit coupling (SOC) at the 2D interface. This effect could indeed be strong at the EuO/graphene interface, forcing and stabilizing an out-off plane magnetization, well above the expected Curie temperature of ferromagnetic layer. However, this would not be consistent with our observation of relatively similar *M(H)* curves in-plane and out-off plane in the SP phase (Fig. 8(e)): by definition, SOC at the graphene interface should lead to highly anisotropic effects inducing large differences in *M* as a function of the direction of the applied magnetic field. We therefore put for another hypothesis: graphene probably modifies, on a given characteristic length, the EuO properties - not only at the very interface with the oxide- probably through bi-axial strain, and could therefore drastically affect the polarons' energy [48] and their temperature-dependent behaviour. As shown on Fig. 4(b), thicker films with *$t_{Eu2O3}$ =14 nm* do not exhibit the high temperature SP phase; the strain due to the graphene layer is probably too low above the characteristic strain-relaxation length and the magnetic behaviour of thicker layers is thus closer to what is observed in bulk EuO.

This hypothesis, related to large strain in the whole EuO layer, does not exclude that graphene could have a specific influence on the EuO SP phase at the very interface: indeed, there is a systematic difference in magnetic polaron sizes as evaluated through AHE or SQUID techniques in our samples, since SQUID gives much larger polarons magnetic dipoles (about one order of magnitude). The most likely explanation is that SQUID observations probe the whole EuO magnetization, whereas AHE is only sensitive to the EuO/Gr interface. This implies that smaller polarons – more precisely, polarons with smaller *m,* maybe due to larger spin disorder- are located at the bottom interface with graphene whereas larger ones are located upper in the oxide layer, but within the characteristic strain-relaxation length. Within such a hypothesis, the SQUID magnetic signal would be dominated by magnetic polarons that graphene cannot feel. The interfacial mechanism that influences the polarons' magnetic dipole *m* at the interface has to be further investigated.

## Conclusion

As a conclusion, we showed that an EuO layer on top of graphene can be obtained by a topotactic reaction with a titanium capping layer. At a given capping temperature of 450°C, the thickness of the reduced EuO layer is limited to about *7 nm.* By adjusting the initial europium oxide film below this threshold thickness, we obtain polycrystalline EuO at the graphene interface which causes a high AHE effect and an original magneto-resistive effect in graphene, both related to the direction of the EuO magnetization and not directly to the applied magnetic field. Moreover, the obtained EuO layer magnetization shows an original behaviour above *$T_c$* in comparison with bulk EuO: a superparamagnetic phase can be promoted in these thin samples above the Curie temperature up to at least 350K. This high temperature magnetism is attributed to a robust magnetic-polaron phase, which was also observed in the bulk EuO, but at much lower temperature. Very interestingly, this magnetic phase also induces proximity Hall effect in graphene. Developing an easy way of depositing EuO on graphene, without need of distillation process in devoted MBE set up, and pushing the magnetism of EuO up to room temperature, should help circumventing the main difficulties for exploiting the spin-polarization of graphene by EuO. The magneto-transport consequences of such an original phase, and the theoretical understanding of the spin-polarization properties of proximized graphene, will require mutual efforts from a wider community of researchers, and will be explored in forthcoming articles.

## VI. Acknowledgements

We thank the STnano clean room facility and the XRD platform of IPCMS for technical support as well as the Labex NIE for partial support. Some of the microscopy works have been conducted in the Laboratorio de Microscopias Avanzadas (LMA) at Universidad de Zaragoza. RA and SH acknowledge funding from the European Union's Horizon 2020 research and innovation program under the Marie Sklodowska-Curie grant agreement number 889546, by the Spanish MCIN (PID2019-104739GB-100/AEI/10.13039/501100011033) and from the European Union H2020 programs "ESTEEM3" (Grant number 823717) and "Graphene Flagship" CORE 3 (Grant number 881603). This study was supported by French state funds managed by the ANR more specifically within the grants MixDferro (ANR-21-CE09-0029) and 2DSwitch (ANR-21-CE09-0031). JFD also acknowledge Junior grant from Institut Universitaire de France.

## VII. References

[1] *Highly efficient spin transport in epitaxial graphene on SiC*, B. Dlubak, M.B. Martin, C. Deranlot, B. Servet, S. Xavier, R. Mattana, M. Sprinkle, C. Berger, W.A. De Heer, F. Petroff, A. Anane, P. Seneor and A. Fert, Nat. Phys. **8**, 557 (2012)

[2] *Electronic spin transport and spin precession in single graphene layers at room temperature,* N. Tombros, C. Jozsa, M. Popinciuc, H.T. Jonkman and B. J. van Wees, Nature **448**, 571 (2007)

[3] *Voltage-controlled inversion of tunnel magnetoresistance in epitaxial nickel/graphene/MgO/cobalt junctions*, F. Godel, M. Venkata Kamalakar, B. Doudin, Y. Henry, D. Halley, and J.-F. Dayen, Appl. Phys. Lett. **105**, 152407 (2014)

[4] *Spin transport in proximity-induced ferromagnetic graphene*, H. Haugen, D Huertas-Hernando and A. Brataas, Phys. Rev. B, **77**, 115406 (2008)

[5] *Spin Lifetimes Exceeding 12 ns in Graphene Nonlocal Spin Valve Devices,* M. Drögeler, C. Franzen, F. Volmer, T. Pohlmann, L. Banszerus, M. Wolter, K. Watanabe, T. Taniguchi, C. Stampfer, and B. Beschoten.

[6] *Ultimate Spin Currents in Commercial Chemical Vapor Deposited Graphene,* J Panda, M Ramu, O Karis, T Sarkar, MV Kamalakar, ACS Nano **14**, 10, 12771 (2020)

[7] *Are $Al_2O_3$ and MgO tunnel barriers suitable for spin injection in graphene?* B. Dlubak, P. Seneor, A. Anane, C. Barraud, C. Deranlot, D. Deneuve, B. Servet, R. Mattana, F. Petroff, and A. Fert Appl. Phys. Lett. **97**, 092502 (2010)

[8] *Two-dimensional van der Waals spinterfaces and magnetic-interfaces*. J.-F. Dayen, S.J. Ray, O. Karis, I.J. Vera-Marun and M.V. Kamalakar, Applied Physics Reviews **7**, 011303 (2020)

[9] *Van der Waals heterostructures for spintronics and opto-spintronics.* J.F. Sierra, J. Fabian, R. Kawakami, S. Roche, S. Valenzuela. *Nat. Nanotechnol.* **16**, 856–868 (2021).

[10] *Proximity Effects Induced in Graphene by Magnetic Insulators: First-Principles Calculations on Spin Filtering and Exchange-Splitting Gaps,* H. X. Yang, A. Hallal, D. Terrade, X. Waintal, S. Roche, and M. Chshiev, Phys. Rev. Lett. 110, 046603 (2013)

[11] *Approaching quantum anomalous Hall effect in proximity-coupled YIG/graphene/h-BN sandwich structure*, C.Tang, B. Cheng, M. Aldosary, Z. Wang, Z. Jiang, K. Watanabe, T. Taniguchi, Marc Bockrath and J. Shi, Appl. Phys. Lett. Mat. **6**, 026401 (2018)

[12] *Proximity-Induced Ferromagnetism in Graphene Revealed by the Anomalous Hall Effect,* Z. Wang, R. Sachs, Y. Barlas and J. Shi, Phys. Rev. Lett., **114,** 016603, (2015)

[13] *Proximity induced room temperature ferromagnetism in graphene probed with spin currents,* J.C. Leutenantsmeyer, A.A Kaverzin, M. Wojtaszek and B.J van Wees, *2D Mater.* **4,** 014001 (2017)

[14] *Strong interfacial exchange field in the graphene/EuS heterostructure* P. Wei, S.Lee, F. Lemaitre, L. Pinel, D. Cutaia, W. Cha, F. Katmis, Y. Zhu, D Heiman, J. Hone, J. S. Moodera and C.T Chen, Nature Mat., **15**,711, (2016)

[15] *Integration of the Ferromagnetic Insulator EuO onto Grahene*, A.G. Swartz, P. M. Odenthal, Y. Hao, R. S. Ruoff, and R.K. Kawakami, ACS Nano **11**, 10063 (2012)

[16] *Structure and Magnetic Properties of Ultra-Thin Textured EuO Films on Graphene*, J. Klinkhammer, D.F. Förster, S. Schumacher, H.P. Oepen, T. Michely, C. Busse, Appl. Phys. Lett., **103**, 131601 (2013)

[17] *Nanoparticle-Induced Anomalous Hall Effect in Graphene*, G. Song, M. Ranjbar, D.R. Daughton,and R.A. Kiehl, Nano Lett. 2019, **19**, 7112 (2019)

[18] *Proximity magnetoresistance in graphene induced by magnetic insulators,* D. A. Solis, A. Hallal, X. Waintal, and M. Chshiev Phys. Rev. B **100**, 104402 (2019)

[19] *Hall Effect in Ferromagnetics*, R. Karplus and J.M Luttinger, Phys. Rev., **95**, 1154 (1954)

[20] *Ferromagnetic Interaction in EuO*, B. T. Matthias, R. M. Bozorth, and J. H. Van Vleck, Phys. Rev. Lett. **7**, 160 (1961)

[21] *Preparation, heat capacity, magnetic properties, and the magnetocaloric effect of EuO,* K. Ahn, J. of Appl. Phys., **97**, 063901 (2005)

[22] *Lower oxides of samarium and europium. The preparation and crystal structure of $SmO_{0.4-0.6}$, SmO and EuO*, H.A. Eick, N.C Baenziger and L.J. Eyring, Am. Chem. Soc. **78**, 5147 (1956).

[23] *High-quality EuO thin films the easy way via topotactic transformation*, T. Mairoser, J.A. Mundy, A. Melville , D. Hodash, P. Cueva, R. Held , A. Glavic, J. Schubert, D.A. Muller , D.G. Schlom and A.Schmehl, Nature Comm. **6**, 7716 (2015)

[24] *Die Gitterkonstanten der C-Formen der Oxyde der seltenen Erdmetalle*, H.Z. Bommer, Z. Anorg. Allg. Chem. **241**, 273 (1939).

[25] *Anisotropic exchange interaction between nonmagnetic europium cations in $Eu_2O_3$*, G. Concas, J. K. Dewhurst, A. Sanna, S. Sharma and S. Massidda, Phys.Rev. B **84**, 014427 (2011)

[26] *High-Temperature Magnetism in Graphene Induced by Proximity to EuO,* D.V. Averyanov, I.S. Sokolov, A. M. Tokmachev, O.E. Parfenov, I.A. Karateev, A. N. Taldenkov, and V.G. Storchak, ACS Appl. Mater. Interfaces **10**, 20767 (2018)

[27] *Experimental advances in charge and spin transport in chemical vapor deposited graphene,* H Mishra, J Panda, M Ramu, T Sarkar, J-F Dayen, Daria Belotcerkovtceva and M Venkata Kamalakar, J. Phys. Mater. **4** 042007 (2021)

[28] *Raman Spectroscopy Study of Rotated Double-Layer Graphene: Misorientation-Angle Dependence of Electronic Structure,* K. Kim, S. Coh, L.Z. Tan, W. Regan, J. Min Yuk, E. Chatterjee, M. F. Crommie, M.L. Cohen, S. G. Louie, and A. Zettl,. Phys. Rev. Lett., **108**, 246103 (2012)

[29] *International Interlaboratory Comparison of Raman Spectroscopic Analysix of CVD-grown Graphene,* P.Turner, K.R Paton, E. Legge, A. de Luna Bugallo *et al*, 2D Materials **9**, 035010 (2022).

[30] *Growth and Characterisation Studies of $Eu_3O_4$ Thin Films Grown on $Si/SiO_2$ and Graphene,* R.O.M Aboljadayel, A. Ionescu, O.J. Burton, G. Cheglakov, S. Hoffmann and C.H.W Barnes, Nanomaterials, **11**, 1598 (2021)

[31] *Electrical Readout of Light-Induced Spin Transition in Thin Film Spin Crossover/Graphene Heterostructure,* N. Konstantinov, A. Tauzin, U. Noumbe, D. Dragoe, B. Kundys, H. Majjad, A. Brosseau, M. Lenertz, A. Singh, S. Berciaud, M.-L. Boillot, B. Doudin, T. Mallahand J.-F. Dayen, J. Mater. Chem. C, **9**, 2712 (2021)

[32] *Room temperature optoelectronic devices operating with spin crossover nanoparticles*, J.-F. Dayen, N. Konstantinov, M. Palluel, N. Daro, B. Kundys, M. Soliman, G. Chastanet and B. Doudin, Mater. Horiz., **8**, 2310 (2021)

[33] *Influence of epitaxial strain on the ferromagnetic semiconductor EuO: First-principles calculations*, N. J. C. Ingle and I. S. Elfimov, Phys. Rev. B **77**, 121202R (2008)

[34] *A magnetic polaron model for the enhanced Curie temperature of $EuO_{1-x}$,* P. Liu and J. Tang, J. Phys. Cond. Matter. **25,** 125802 (2013)

[35] *Exploring the intrinsic limit of the charge-carrier-induced increase of the Curie temperature of Lu- and La-doped EuO thin films*, R. Held, T. Mairoser, A. Melville, J. A. Mundy, M. E. Holtz, D. Hodash, Z. Wang, J. T. Heron, S. T. Dacek, B. Holländer, D. A. Muller, and D. G. Schlom, Phys. Rev. Mat. **4**, 104412 (2020)

[36] *Fine structure of metal–insulator transition in EuO resolved by doping engineering*, D.V Averyanov, O.E Parfenov, A.M Tokmachev, I.A Karateev, O.A Kondratev, A.N Taldenkov, M.S Platunov, F. Wilhelm, A. Rogalev and V.G Storchak, Nanotechnology **29**, 195706 (2018)

[37] *Epitaxial integration of the highly spin-polarized ferromagnetic semiconductor EuO with silicon and GaN*, A. Schmelh, V. Vaithyanathan, A. Herrnberger, S. Thiel, C. Richter *et al.*, Nat. Mat., **6**, 882 (2007)

[38] *Magnetic Ordering in $Eu_3O_4$ and $EuGd_2O_4$*, L. Holmes and M. Schieber, J. of Appl. Phys. **37**, 968 (1966)

[39] *Electronic structure of a mixed valence system: $Eu_3O_4$*, B. Batlogg, E. Kaldis, A. Schlegel and P. Wachter, Phys. Rev. B **12**, 3940 (1975)

[40] *Anomalous anisotropic magnetoresistance effects in graphene*, Y. Liu, R. Yang, H. Yang, D. Wang, Q. Zhan, G. Zhang, Y. Xie, B. Chen and R.W Lei, AIP Adv. **4**, 097101 (2014)

[41] *Strong Suppression of Weak Localization in Graphene,* S. V. Morozov, K. S. Novoselov, M. I. Katsnelson, F. Schedin, L. A. Ponomarenko, D. Jiang, and A. K. Geim, Phys. Rev. Lett. **97**, 016801 (2006)

[42] *2D ferromagnetism in europium/graphene bilayers*, I.Sokolov, D.V. Averyanov, O.E. Parfenov, I.A. Karateev, A.N, Taldenkov, A.M. Tokmachev and V.G. Storchak, Mater. Horiz., **7**, 1372 (2020)

[43] *Magnetotransport signatures of the proximity exchange and spin-orbit coupling in graphene*, J. Lee and J. Fabian, Phys. Rev. B., **94**, 195401 (2016)

[44] *Magnetic polarons and the metal-semiconductor transitions in $(Eu,La)B_6$ and EuO: Raman scattering studies*, , C. S. 45 and S. L. Cooper D. P. Young, Z. Fisk, A. Comment and J.P. Ansermet, Phys. Rev. B, **64**, 174412 (2001)

[45] *Hypergiant spin polarons photogenerated inferromagnetic europium chalcogenides*, X. Gratens, Yunbo Ou, J. S. Moodera, P. H. O. Rappl, and A. B. Henrique, Appl. Phys. Lett. **116**, 152402 (2020)

[46] *A high-temperature ferromagnetic topological insulating phase by proximity coupling,* F. Katmis, V. Lauter, F.S. Nogueira, B.A. Assaf, M.E. Jamer, P. Wei, B. Satpati, J.W. Freeland, I. Eremin, D.Heiman, P. Jarillo-Herrero and J. S. Moodera, Nature **533**, 513 (2016)

[47] *Crossover from lattice to plasmonic polarons of a spin-polarised electron gas in ferromagnetic EuO,* J.M. Riley, F. Caruso, C. Verdi, L.B. Duffy, M.D. Watson, L. Bawden, K. Volckaert, G. van der Laan, T. Hesjedal, M. Hoesch, F. Giustino and P.D.C. King, Nature Comm.,**9**, 2305 (2018)